\documentstyle[12pt,epsfig]{article}

\newcommand{\beq}{\begin{equation}}
\newcommand{\eeq}{\end{equation}}
\newcommand{\barr}{\begin{eqnarray}}
\newcommand{\earr}{\end{eqnarray}}

\newcommand{\andy}[1]{ }

\def\As{{\cal A}}

\newcommand{\bm}[1]{\mbox{\boldmath $#1$}}

\newcommand{\ket}[1]{| #1 \rangle}
\newcommand{\bra}[1]{\langle #1 |}

\begin{document}

\begin{titlepage}
\begin{flushright}
\today \\
BA-TH/99-366 \\
\end{flushright}
\vspace{.5cm}
\begin{center}
{\LARGE Van Hove's ``$\lambda^2t$" limit in nonrelativistic and
relativistic field-theoretical models

}

\quad

{\large P. FACCHI\footnote{email: Paolo.Facchi@ba.infn.it}
        and S. PASCAZIO\footnote{email: Saverio.Pascazio@ba.infn.it}\\
           \quad    \\

        Dipartimento di Fisica, Universit\`a di Bari

        and Istituto Nazionale di Fisica Nucleare, Sezione di Bari \\
 I-70126  Bari, Italy \\

}

\vspace*{.5cm} PACS numbers: 05.40.-a; 31.70.Hq, 03.65.-w; 31.30.Jv
\vspace*{.5cm}

{\small\bf Abstract}\\ \end{center}

{\small Van Hove's ``$\lambda^2 t$" limiting procedure is analyzed in
some interesting quantum field theoretical cases, both in
nonrelativistic and relativistic models. We look at the deviations
from a purely exponential behavior in a decay process and discuss the
subtle issues of state preparation and initial time.}

\vspace*{1cm} Keywords: Van Hove's limit, Exponential Law, Fermi
Golden Rule

\end{titlepage}

\newpage

\setcounter{equation}{0}
\section{Introduction  }
\label{sec-introd}
\andy{intro}

In 1955 Van Hove proposed a remarkable time rescaling procedure
\cite{vanHove} that enabled him to derive the master equation
from the Schr\"odinger equation, for a quantum mechanical system
endowed with an infinite number of degrees of freedom, such as a
quantum field. The idea is to consider the limit
\andy{vHlimit}
\beq
\lambda\to0 \quad\mbox{keeping}\quad \tilde t =
\lambda^2t\quad\mbox{finite ($\lambda$-independent constant)},
\label{eq:vHlimit}
\eeq
where $\lambda$ is the coupling constant and $t$ time. One then looks
at the evolution of the quantum system as a function of the rescaled
time $\tilde t$. This is called Van Hove's ``$\lambda^2 t$" limit and
provided an interesting solution to some long-standing problems in
quantum mechanics and quantum field theory, such as a rigorous
justification of the Fermi ``golden" rule \cite{Fermigold} and of the
Weisskopf-Wigner approximation \cite{seminal}.

Van Hove's prescription avoided the rigorous consequences of the
quantum mechanical evolution law, which is governed by strictly
unitary operators and predicts that the decay of an unstable quantum
system cannot be purely exponential, being quadratic for very short
times \cite{shortt} and given by a power law for very long times
\cite{longtt}. These features of the quantum evolution are so well known
that they are discussed even in textbooks of quantum mechanics
\cite{Sakurai} and quantum field theory \cite{Brown}. The temporal
behavior of quantum systems is reviewed in Ref.\ \cite{temprevi}.

One should notice that the problem of the deviations from exponential
decay was considered an academic one until very recently. The renewed
interest in the short-time nonexponential behavior was caused by a
nice proposal by Cook \cite{Cook}, the subsequent experiment
performed by Itano {\em et al} \cite{Itano} and the debate that
followed \cite{Itanodiscuss}. One must notice, however, that Cook's
idea and the subsequent papers did not deal with {\em bona fide}
unstable systems. The latter require a quantum field theoretical
analysis and the careful treatment of cut-offs and divergent
quantities
\cite{FP1,BMT,Kurizki}. It is also worth emphasizing that no
deviations from the exponential behavior for an unstable system were
observed until 1997, when Raizen's group detected non-exponential
leakage through a potential barrier
\cite{Raizen}.

In this paper we shall look at Van Hove's limit from the perspective
of the complex energy plane. We shall consider some particular cases,
concentrating our attention on the exponential decay law and the
irreversible features \cite{VanKampen} of the evolution. There is
interesting related work in the literature, in particular in
connection with quantum dynamical semigroups \cite{Bohmbook} and the
so-called ``stochastic limit" in quantum theory
\cite{Accardi}.

\section{A simple example: $N$-level atom }
\label{sec-VHNlev}
\andy{VHNlev}

We start our analysis by considering a simple nonrelativistic model:
an $N$-level atom in interaction with the electromagnetic field
\cite{hydrovanH}.
This example will help us to pin down some salient features of the
$\lambda^2t$ limit. The Hamiltonian is
\andy{newHV}
\beq
H=H_0+\lambda V,
\label{eq:newHV}
\eeq
with ($\hbar=c=1$)
\andy{newH0,V}
\barr
H_0 & \equiv & \sum_\nu \omega_\nu b^\dagger_\nu b_\nu +
\sum_\beta
\int_0^\infty d\omega \,
\omega a^\dagger_{\omega\beta} a_{\omega\beta} , \label{eq:newH0} \\
 V & = & \sum_{\mu,\nu} \sum_\beta \int_0^\infty
d\omega \left[ \varphi^{\mu\nu}_\beta(\omega) b^\dagger_\mu b_{\nu}
a^\dagger_{\omega \beta}
+ \varphi^{\mu\nu*}_\beta(\omega) b^\dagger_{\nu} b_\mu a_{\omega \beta}
\right],
\label{eq:newV}
\earr
where $\nu$ runs over all the atomic states, $b^\dagger_\nu, b_\nu$
are the annihilation and creation operators of the atomic level
$\nu$, obeying anticommutation relations
 \beq
 \{b_k, b^\dagger_\ell \} = \delta_{k\ell},
 \label{eq:fermicomm}
 \eeq
and $a^\dagger_{\omega\beta}, a_{\omega\beta}$ are the annihilation
and creation operators of the electromagnetic field, satisfying
commutation relations
 \andy{boscomm2}
 \beq
 [a_{\omega \beta}, a^\dagger_{\omega' \beta'}] =
\delta(\omega-\omega')\delta_{\beta \beta'},
 \label{eq:boscomm2}
 \eeq
where $\omega$ is energy and $\beta$ stands for other (discrete)
quantum numbers (e.g.\, $\beta = (j,m,\epsilon$), where $j$ is the
total angular momentum (orbital+spin) of the photon, $m$ its magnetic
quantum number and $\epsilon$ defines the photon parity
$P=(-1)^{j+1+\epsilon}$). The general features of the form factors
are well known for a wide class of physical systems \cite{Heitler}
and some particular cases of the above Hamiltonian have been widely
investigated in the literature \cite{KnightMilonni'76}.

\begin{figure}
\epsfig{file=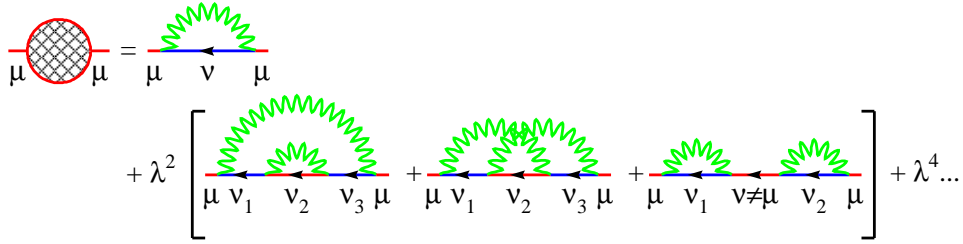, width=13.5cm}
\caption{Graphic representation of (\ref{eq:SEF}):
$\Sigma^{(2)}$ and $\Sigma^{(4)}$ are in the first and second line,
respectively.}
\label{fig:sigma4}
\andy{sigma4}
\end{figure}

Assume one can prepare, say at time $t=0$, the system in the initial
state $|\mu;0\rangle$ (atom in state $\mu$ and no photons). The
problem of state preparation is a subtle one that will be carefully
discussed later. The initial state is an eigenstate of the
unperturbed Hamiltonian $H_0$ and the evolution is governed by the
unitary operator
\andy{survFT}
\beq
\label{eq:survFT}
U(t)= \exp (-iHt) =
\frac{i}{2\pi}\int_{\rm C}dE\frac{e^{-iEt}}{E-H},
\eeq
where the path C is a straight horizontal line just above the real
axis. By defining the resolvents ($\Im E>0$)
\beq
S(E)\equiv\langle
\mu;0|\frac{1}{E-H_0}|\mu;0\rangle=\frac{1}{E-\omega_\mu},
\qquad
S'(E)\equiv\langle \mu;0|\frac{1}{E-H}|\mu;0\rangle,
\eeq
Dyson's resummation reads
\andy{S'S}
\beq
S'(E)=S(E)+\lambda^2S(E)\Sigma(E)S(E)+\lambda^4S(E)\Sigma(E)S(E)
\Sigma(E)S(E)+\dots ,
\label{eq:S'S}
\eeq
where $\Sigma(E) = \langle \mu;0|V(E-H_0)^{-1}V|\mu;0\rangle$ is the
1-particle irreducible self-energy function, that can be evaluated by
the expansion
\andy{SEF}
\beq
\Sigma(E)=\Sigma^{(2)}(E)+\lambda^2\Sigma^{(4)}(E)+\dots,
\label{eq:SEF}
\eeq
with
\andy{gensigma2}
\beq\label{eq:gensigma2}
\Sigma^{(2)}(E)\equiv\sum_{\nu,\beta}\int_0^\infty d\omega
\frac{|\varphi^{\nu\mu}_\beta(\omega)|^2}{E-\omega_\nu-\omega} .
\eeq
Both $\Sigma^{(2)}$ and $\Sigma^{(4)}$ are shown as Feynman diagrams
in Figure
\ref{fig:sigma4}.
In the complex $E$-plane $\Sigma(E)$ and $\Sigma^{(2)}(E)$ have a
branch cut running from the ground-state energy to $\infty$ and no
singularity on the first Riemann sheet. Summing the series
(\ref{eq:S'S}) one obtains
\andy{sumser}
\beq
S'(E)=\frac{1}{S(E)^{-1}-\lambda^2\Sigma(E)}=
\frac{1}{E-\omega_\mu-\lambda^2\Sigma(E)} .
\label{eq:sumser}
\eeq
We define the ``survival" or nondecay amplitude and probability at
time $t$ (interaction picture)
 \andy{survampl,ndq1}
 \barr
 \As(t) & = & \langle \mu;0|e^{i H_0 t}U(t)|\mu;0\rangle ,
\label{eq:survampl} \\
P(t) & = & |\langle \mu;0 |e^{i H_0 t}U(t) |\mu;0 \rangle |^2.
\label{eq:ndq1}
\earr
Incidentally, notice that the survival probability at short times
behaves quadratically
\andy{naiedef}
\beq
P(t) = 1 - t^2/\tau_{\rm Z}^2 + \cdots,
\qquad \tau_{\rm Z} \equiv (\lambda^2\langle \mu;0|V^2|\mu;0\rangle)^{-1/2} .
\label{eq:naiedef}
\eeq
The quantity $\tau_{\rm Z}$ is the ``Zeno time:" it is the convexity
of $P(t)$ in the origin. The nonexponential behavior at short times,
besides its fundamental interest, entails the quantum Zeno effect
\cite{shortt}. Notice that the expansion (\ref{eq:naiedef}) is
formal: we are implicitly requiring that the second moment of the
interaction Hamiltonian exists---a delicate assumption in quantum
field theory \cite{FP1,BMT}.

The survival amplitude can be expressed as
\andy{survE}
\beq
\As(t)=
\frac{i}{2\pi}\int_{\rm C}dE e^{-iEt} S'(E+\omega_0)
     = \frac{i}{2\pi}\int_{\rm C}dE\frac{e^{-iEt}}
{E-\lambda^2\Sigma(E+\omega_0)} .
\label{eq:survE}
\eeq
In Van Hove's limit one looks at the evolution of the system over
time intervals of order $t=\tilde t/\lambda^2$ ($\tilde t$
independent of $\lambda$), in the limit of small $\lambda$. Let us
see how this procedure works in the complex-energy plane. To this
end, by rescaling time $\tilde t\equiv \lambda^2 t$, we  can write
\andy{survEsc}
\beq
\As\left(\frac{\tilde{t}}{\lambda^2}\right)=
\frac{i}{2\pi}\int_{\rm C}
d\tilde E\frac{e^{-i\tilde{E}\tilde t}} {\tilde
E-\Sigma(\lambda^2\tilde E+\omega_0)},
\label{eq:survEsc}
\eeq
where we are naturally led to introduce the rescaled energy
$\widetilde E\equiv E/\lambda^2$. Taking the Van Hove limit we get
\andy{sigma2lim}
\barr
\Sigma(\lambda^2\tilde E+\omega_\mu)&\stackrel{\lambda\rightarrow 0}
{\longrightarrow}&\Sigma^{(2)}(\lambda^2\tilde E+\omega_\mu)\Big|_
{\lambda=0}=\Sigma^{(2)}(\omega_\mu+i0^+) \nonumber \\
&=& \sum_{\nu,\beta}\int_0^\infty d\omega
\frac{|\varphi^{\nu\mu}_\beta(\omega)|^2}{\omega_\mu-\omega_\nu-\omega+i0^+}
\equiv \Delta(\omega_\mu)-\frac{i}{2}\Gamma(\omega_\mu) ,  \nonumber \\
\label{eq:sigma2lim}
\earr
where
\andy{polecoord1,2}
\barr
\Delta(\omega_\mu)&\equiv& {\cal P} \sum_{\nu,\beta}\int_0^\infty d\omega
\frac{|\varphi^{\nu\mu}_\beta(\omega)|^2}{\omega_\mu-\omega_\nu-\omega+i0^+},
\label{eq:polecoord1} \\
\Gamma(\omega_\mu)&\equiv& 2\pi \sum_{\nu,\beta} |\varphi^{\nu\mu}_\beta(\omega)|^2,
\label{eq:polecoord2}
\earr
the term $+i0^+$ being due to the fact that $\Im\tilde E>0$. The
propagator becomes
\beq
\widetilde S'(\tilde E)=\lim_{\lambda\to0}
\frac{1}{\tilde E-\Sigma^{(2)}
(\lambda^2\tilde E+\omega_\mu) +{\rm O}(\lambda^2)}
=\frac{1}{\tilde E-\Sigma^{(2)}(\omega_\mu+i0^+)}
\eeq
\begin{figure}
\epsfig{file=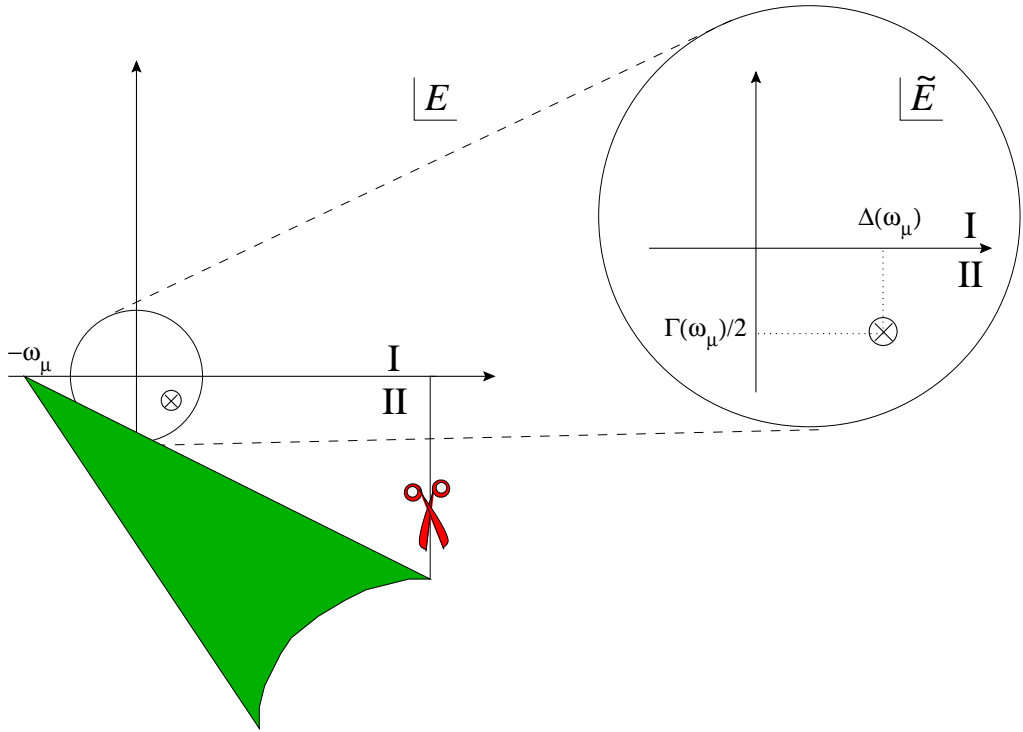, width=13.5cm}
\caption{Singularities of the propagator (\ref{eq:survE}) in the complex-$E$
plane. The first Riemann sheet ({\rm I}) is singularity free. The
logarithmic cut is due to $\Sigma^{(2)}(E)$ and the pole is located
on the second Riemann sheet ({\rm II}). The Van Hove rescaling
procedure acts as a ``magnifying glass" in the complex energy plane.
After rescaling, the pole has coordinates
(\ref{eq:polecoord1})-(\ref{eq:polecoord2}) in the complex-$\tilde E$
plane, without higher-order corrections in the coupling constant. }
\label{fig:cmplxe}
\andy{sigma4}
\end{figure}
and the survival probability reads
\andy{survEsclim}
\beq
\widetilde \As(\tilde{t})\equiv
\lim_{\lambda\to0}\As\left(\frac{\tilde{t}}{\lambda^2}\right)
=\frac{i}{2\pi}\int_{\rm C} d\tilde E e^{-i\tilde{E}\tilde
t}\widetilde S'(\tilde E)= e^{-\left[i
\Delta(\omega_\mu)+\Gamma(\omega_\mu)/2\right]t},
\label{eq:survEsclim}
\eeq
which yields a purely exponential decay (Weisskopf-Wigner
approximation and Fermi golden rule). In Figure \ref{fig:cmplxe} we
endeavoured to clarify the role played by the time-energy rescaling
in the complex-$E$ plane.

A few comments are in order. In the present model, the Van Hove limit
works in two ``steps." First, it constrains the evolution in a
Tamm-Dancoff sector \cite{TammDancoff}: the system can only
``explore" those states that are directly related to the initial
state $\mu$ by the interaction $V$: the ``excitation number" ${\cal
N}_\mu \equiv b^\dagger_\mu b_\mu + \sum_{\beta,
\omega} a^\dagger_{\omega\beta} a_{\omega\beta}$ becomes a conserved
quantity and, as a consequence, the self-energy function consists
only of a second order contribution that can be evaluated exactly.
Second, it reduces this second order contribution, which depends on
energy as in (\ref{eq:gensigma2}), to a constant (its value in the
energy $\omega_\mu$ of the initial state), like in
(\ref{eq:sigma2lim}). Hence the analytical properties of the
propagator, which had branch-cut singularities, reduce to those of a
single complex pole, whose imaginary part (responsible for
exponential decay) yields the Fermi golden rule, evaluated at second
order of perturbation theory.

Notice that it is the latter step (and not the former one) which is
strictly necessary to obtain a dissipative behavior: Indeed,
substitution of the pole value in the total self-energy function
yields exponential decay, including, as is well known, higher-order
corrections to the Fermi golden rule. On the other hand, the first
step is very important when one is interested in computing the
leading order corrections to the exponential behavior. To this
purpose one can solve the problem in a restricted Tamm-Duncoff sector
of the total Hilbert space (i.e., in an eigenspace of ${\cal N}_\mu$
--- in our case, ${\cal N}_\mu=1$) and exactly evaluate the evolution
of the system with its deviations from exponential law.

\setcounter{equation}{0}
\section{A more general framework}
\label{sec-general}
\andy{general}
Let us generalize the analysis of the previous section. Consider the
Hamiltonian
\beq
H=H_0+\lambda V
\eeq
and suppose that one can prepare an initial state $|a\rangle$ with
the following properties
\andy{aprop}
\barr
& & H_0|a\rangle=E_a|a\rangle,\qquad
\langle a|V|a\rangle=0, \nonumber\\
& & \langle a|a\rangle=1.
\label{eq:aprop}
\earr
The survival amplitude of state $|a\rangle$ reads
\andy{survEgen}
\barr
\As(t)&\equiv&\langle a|e^{i H_0 t} U(t)|a\rangle=
\frac{i}{2\pi}\int_{\rm C}dE e^{-iEt} S'(E+E_a)\nonumber\\
&=&
\frac{i}{2\pi}\int_{\rm C}dE\frac{e^{-iEt}}
{E-\lambda^2\Sigma(E+E_a)},
\label{eq:survEgen}
\earr
where $S'(E)\equiv\langle a|(E-H)^{-1}|a\rangle$ and $\Sigma(E)$ is
the 1-particle irreducible self-energy function, that can be
expressed by a perturbation expansion
\andy{sigmagen}
\beq
\lambda^2 \Sigma(E) = \lambda^2 \Sigma^{(2)}(E) + \lambda^4 \Sigma^{(4)}(E)
+\cdots .
\label{eq:sigmagen}
\eeq
The second order contribution has the general form
\andy{sigma2gen}
\barr
\Sigma^{(2)}(E)&\equiv&\langle a|V P_d \frac{1}{E-H_0} P_d V|a\rangle
=\sum_{n\neq a} \left| \langle a|V|n\rangle \right|^2\frac{1}{E-E_n}
\nonumber\\
&=& \int_0^\infty \frac{dE'}{2\pi} \frac{\Gamma(E')}{E-E'},
\label{eq:sigma2gen}
\earr
where $P_d=1-|a\rangle\langle a|$ is the projector over the decayed
states, $\{|n\rangle\}$ is a complete set of eingenstates of $H_0$
($H_0|n\rangle=E_n|n\rangle$ and we set $E_0=0$) and
\andy{genGamma}
\beq
\Gamma(E)\equiv2\pi\sum_{n\neq a}\left| \langle a|V|n\rangle \right|^2\delta(E-E_n).
\label{eq:genGamma}
\eeq
Notice that $\Gamma(E)\geq0$ for $E>0$ and is zero otherwise. In
the Van Hove limit we  get
\andy{gensurvEsclim}
\beq
\widetilde \As(\tilde{t})\equiv
\lim_{\lambda\to0}\As\left(\frac{\tilde{t}}{\lambda^2}\right)
=\frac{i}{2\pi}\int_{\rm C}
d\tilde E e^{-i\tilde{E}\tilde t}\widetilde S'(\tilde E),
\label{eq:gensurvEsclim}
\eeq
where the resulting propagator in the rescaled energy
$\tilde E=E/\lambda^2$ reads
\andy{genrescS}
\beq
\widetilde S'(\tilde E)
=\frac{1}{\tilde E-\Sigma^{(2)}(E_a+i0^+)},
\label{eq:genrescS}
\eeq
where we used
\andy{gensigma2lim}
\beq
\Sigma(\lambda^2\tilde E+E_a)\stackrel{\lambda\rightarrow 0}
{\longrightarrow}\Sigma^{(2)}(\lambda^2\tilde E+E_a)\Big|_
{\lambda=0}=\Sigma^{(2)}(E_a+i0^+)
\label{eq:gensigma2lim}
\eeq
(Weisskopf-Wigner approximation and Fermi golden rule).

Let us compute the leading order corrections to the exponential
behavior, in particular at short times. Just above the positive real
axis we can write
\andy{sigmaEa}
\beq
\label{eq:sigmaEa}
\Sigma^{(2)}(E+i0^+)=\Delta(E)-\frac{i}{2}\Gamma(E),
\eeq
where
\andy{genpolecoord1}
\beq
\Delta(E) = {\cal P}\int_0^\infty \frac{dE'}{2\pi}\frac{\Gamma(E')}{E-E'}.
\label{eq:genpolecoord1} \\
\eeq
We assume that $\Gamma(E)$ is sommable in $(0,+\infty)$, so that for
some $\eta>0$,
\andy{Gammalim}
\beq
\Gamma(E)\propto E^{\eta-1}\quad\mbox{for}\quad E\to0.
\label{eq:Gammalim}
\eeq
It is then straightforward to obtain
 \andy{gentauZ,genfgr}
 \barr
 \tau_{\rm Z} &=& \frac{1}{\lambda} \left[\int_0^\infty
 \frac{dE}{2\pi}\Gamma(E)\right]^{-1/2},
 \label{eq:gentauZ}\\
 \tau_{\rm E} &=& \frac{1}{\lambda^2 \Gamma(E_a)},
 \label{eq:genfgr}
\earr
which are the Zeno time and the lifetime, respectively. When time is
rescaled  according to Van Hove, the Zeno region vanishes
\andy{resctauZ}
\beq
\tilde\tau_{\rm Z}\equiv\lambda^2\tau_{\rm Z}=\lambda\left[\int_0^\infty
 \frac{dE}{2\pi}\Gamma(E)\right]^{-1/2}={\rm O}(\lambda)
\label{eq:resctauZ}
\eeq
and the lifetime reads
\andy{resctauE}
\beq
\tilde\tau_{\rm E}\equiv \lambda^2\tau_{\rm E}=\frac{1}{\Gamma(E_a)}
.
\label{eq:resctaupow}
\eeq
It goes without saying that the evolution must then be described in
terms of the rescaled time $\tilde t=\lambda^2 t$.

\begin{figure}
\epsfig{file=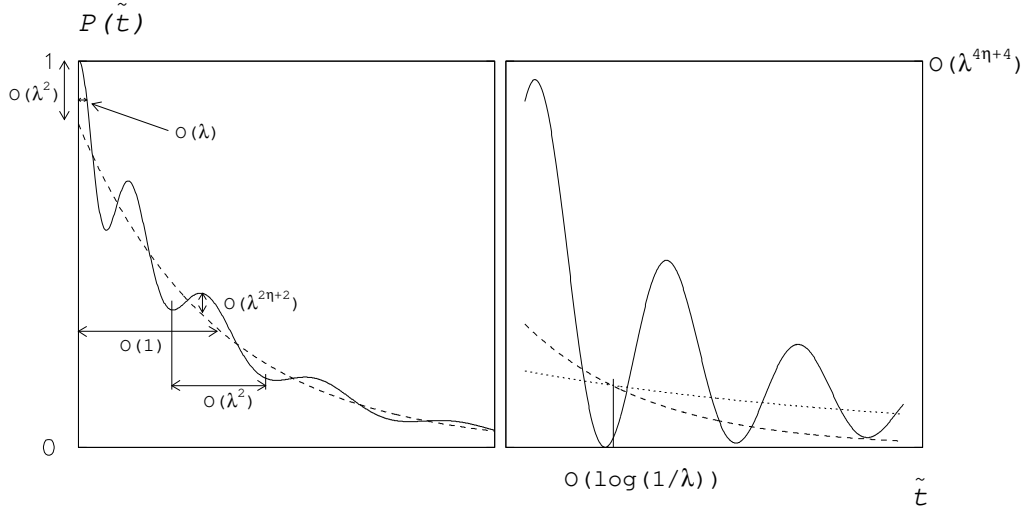, width=13.5cm}
\caption{Essential features (not in scale!) of the survival probability
as a function of the rescaled time $\tilde t$. The Zeno time is ${\rm
O}(\lambda)$, the lifetime ${\rm O}(1)$, during the whole evolution
there are oscillations of amplitude ${\rm O}(\lambda^{2\eta+2})$ and
the transition to a power law occurs after a time ${\rm O}(\log
(1/\lambda))$ [see (\ref{eq:resctauZ})-(\ref{eq:resctaupow})]. The
normalization factor becomes unity like $1-{\rm O}(\lambda^2)$. The
dashed line is the exponential and the dotted line the power law.}
\label{fig:summa}
\andy{summa}
\end{figure}

The details of the evolution were thoroughly investigated in
\cite{hydrovanH} in terms of the coupling constant.
We only show in Figure \ref{fig:summa} the most salient features of
the survival probability.
In the Van Hove limit the coupling constant vanish ($\lambda \to 0$)
and several things happen at once: the initial quadratic (quantum
Zeno) region vanish, the oscillations are ``squeezed" out and the
power law is ``pushed" to infinity: only a clean exponential law is
left at all times, with the right normalization factors. All this is
not surprising, being implied by the Weisskopf-Wigner approximation.
However, the concomitance of these features is so remarkable that one
cannot but wonder at the effectiveness of this limiting procedure.

In atomic and molecular physics one gets very small deviations from
the exponential law. For this reason, we displayed in Figure
\ref{fig:summa} the survival probability  by greatly exaggerating its
most salient features.

\setcounter{equation}{0}
\section{Relativistic quantum field theory}
\label{sec-relat}
\andy{relat}

\begin{figure}
\epsfig{file=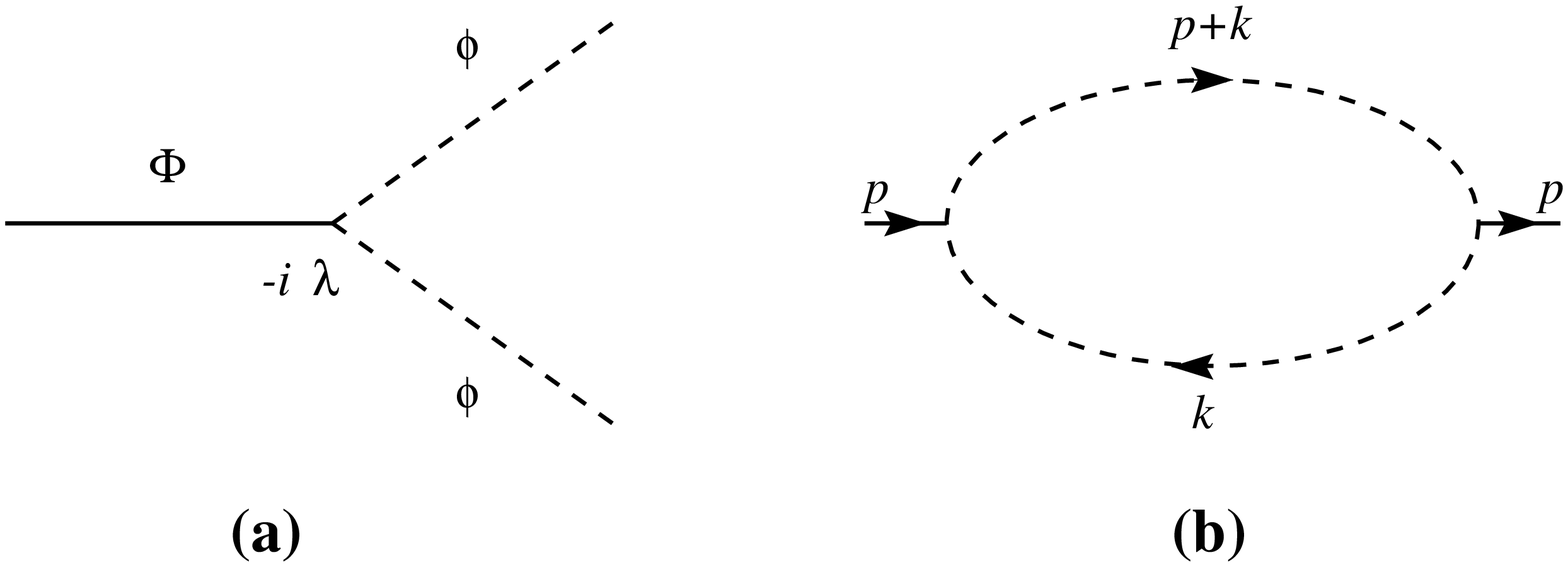, width=13.5cm}
\caption{(a) Decay of $\Phi$ into $\phi+\phi$. (b) Loop contribution to $\Sigma^{(2)}(p^2)$.}
\label{fig:loop}
\andy{loop}
\end{figure}

We look now at a more complicated system. Consider the decay of a
massive scalar particle $\Phi$ of mass $M$ into two identical massive
scalar particles $\phi$ of mass $m$ [See Fig.\ \ref{fig:loop}(a)]. We
shall work in a completely relativistic framework. The Lagrangian
density of our model reads
\beq
{\cal L}=\frac{1}{2}(\partial_\mu\Phi)^2-\frac{1}{2}M^2\Phi^2
+\frac{1}{2}(\partial_\mu\phi)^2-\frac{1}{2}m^2\phi^2
-\frac{\lambda}{2}\mu \Phi\phi^2 + {\cal L}_{\rm CT} .
\eeq
The Lagrangian  ${\cal L}_{\rm CT}$ contains the counterterms
absorbing the infinite but unobservable shifts between the bare
parameters ($M_0$, $m_0$, $\lambda_0$) and the physical ones ($M$,
$m$, $\lambda$):
\beq
{\cal L}_{\rm
CT}=\frac{1}{2}\delta_Z(\partial_\mu\Phi)^2-\frac{1}{2}\delta_M\Phi^2
+\frac{1}{2}\delta_z(\partial_\mu\phi)^2-\frac{1}{2}\delta_m\phi^2
-\frac{\delta_\lambda}{2}\mu \Phi\phi^2 ,
\eeq
with
\andy{deltaz}
\barr
\delta_Z&=&Z-1,\qquad \delta_z=z-1, \nonumber \\
\delta_M&=&M^2_0 Z -M^2, \quad \delta_m=m^2_0 z -m^2, \quad
\delta_\lambda=\lambda_0 z Z^{1/2} - \lambda,
\label{eq:deltaz}
\earr
where $Z$ and $z$ are the field-strength renormalization constants
($\Phi_0=Z^{1/2}\Phi$ and $\phi_0=Z^{1/2}\phi$).

The full two-point function
\andy{corr.func}
\beq
G(p)\equiv\int d^4x\; e^{ip\cdot x}
\bra{\Omega}T\Phi(x)\Phi(0)\ket{\Omega}
\label{eq:corr.func}
\eeq
is given by Dyson's resummation of the geometric series:
\barr
G(p)&=&\frac{i}{p^2-M^2+i0^+}
+\frac{i}{p^2-M^2+i0^+}(-i\Sigma(p^2))\frac{i}{p^2-M^2+i0^+}+\cdots
\nonumber\\
&=&\frac{i}{p^2-M^2-\Sigma(p^2)+i0^+},
\earr
where $\Sigma(p^2)$ is the 1-particle irreducible self-energy:
\beq
\Sigma(p^2)=\lambda^2\Sigma^{(2)}(p^2)+\lambda^4\Sigma^{(4)}(p^2)+\cdots.
\eeq
By using Feynman rules it is straightforward to write down the
contribution of the loop diagram in Fig.~\ref{fig:loop}(b):
\beq
-i\Sigma^{(2)}_{\rm loop}(p^2)\equiv\frac{(-i\mu)^2}{2}
\int
\frac{d^4k}{(2\pi)^4}\frac{i}{k^2-m^2+i0^+}\frac{i}{(p+k)^2-m^2+i0^+}.
\eeq
We compute $\Sigma^{(2)}_{\rm loop}$ by using dimensional
regularization. Introducing a Feynman parameter $\xi$ and shifting
the integration variable ($q=k+xp$) we get
\barr
-i\Sigma^{(2)}_{\rm loop}(p^2)&=&\frac{(-i\mu)^2}{2}
\int_0^1 d\xi \int \frac{d^D k}{(2\pi)^D}
\frac{i^2}{[k^2+2\xi k\cdot p+\xi p^2-m^2]^2}
\nonumber\\
&=&\frac{(-i\mu)^2}{2}
\int_0^1 d\xi \int \frac{d^D q}{(2\pi)^D}
\frac{i^2}{[q^2+\xi(1-\xi)p^2-m^2]^2}.
\earr
Performing a Wick rotation to Euclidean space ($q^0_{\rm E}=-iq^0$)
and evaluating the momentum integral we obtain
\barr
\Sigma^{(2)}_{\rm loop}(p^2)&=&-\frac{\mu^2}{2}
\int_0^1 d\xi \int \frac{d^D q_{\rm E}}{(2\pi)^D}
\frac{1}{[q_{\rm E}^2-\xi(1-\xi)p^2+m^2]^2}\nonumber\\
&=&-\frac{\mu^2}{2(4\pi)^2}
\Gamma\left(2-\frac{D}{2}\right)\int_0^1 d\xi
\left(\frac{m^2-\xi(1-\xi)p^2}{4\pi}\right)^{\frac{D}{2}-2} ,
\earr
which diverges like $2/(4-D)$ for $D\to4$:
\beq
\Sigma^{(2)}_{\rm loop}(p^2)\sim
-\frac{\mu^2}{2(4\pi)^2}\int_0^1 d\xi
\left[\frac{2}{4-D}-\gamma-\log\left(
\frac{m^2-\xi(1-\xi)p^2}{4\pi}\right)\right] ,
\eeq
where $\gamma$ is the Euler-Mascheroni constant. Let us define the
particle mass $M$ by the condition
\beq
{\rm Re}\Sigma(M^2)=0,
\label{eq:ren.cond}
\eeq
so that the propagator has the following behavior
\beq
G(p)\sim\frac{i \bar{Z}}{p^2-M^2-i\bar{Z}{\rm Im}\Sigma(M^2)},
\quad \mbox{for} \quad p^2\to M^2,
\eeq
with $\bar{Z}^{-1}=1-\Sigma'(M^2)$. Notice that $\bar Z$ is a finite
field-strength normalization constant. Imposing the renormalization
condition (\ref{eq:ren.cond}) one gets
\barr
\Sigma^{(2)}(p^2)&=&\Sigma^{(2)}_{\rm loop}(p^2)
-{\rm Re}\Sigma^{(2)}_{\rm loop}(M^2)\nonumber\\
&=&\frac{\mu^2}{2(4\pi)^2}\left[\int_0^1 d\xi
\log\left(\frac{m^2-\xi(1-\xi)p^2}{m^2}\right)-{\cal C}\right],
\earr
where
\beq
{\cal C}\equiv\int_0^1 d\xi
\log\left|\frac{m^2-\xi(1-\xi)M^2}{m^2}\right| .
\eeq
This is equivalent to setting  $\delta_M=-\lambda^2 {\rm
Re}\Sigma^{(2)}_{\rm loop}(M^2)$ and $\delta_Z=0$, in Eq.\
(\ref{eq:deltaz}). The function $\Sigma^{(2)}(s)$ is analytic in the
cut $s$ plane and the discontinuity across the cut is $( x>4m^2 )$
\beq
\Sigma^{(2)}(x+i0^+)-\Sigma^{(2)}(x-i0^+)=
-2\pi i \frac{\mu^2}{2(4\pi)^2} \sqrt{1-\frac{4m^2}{x}} .
\eeq
Hence $\Sigma^{(2)}(s)$ can be represented by a dispersion relation.
It is indeed straightforward to obtain
\beq
\Sigma^{(2)}(s)=
\frac{\mu^2}{2(4\pi)^2} \left[s\int_{4m^2}^\infty ds'
\frac{\rho(s')}{s' (s-s')} - {\cal C}\right],
\eeq
with
\beq
\rho(s)=\sqrt{1-\frac{4m^2}{s}}.
\eeq
Therefore the propagator (\ref{eq:corr.func})
\beq
G(s)=\frac{i}{s-M^2-\Sigma(s)}
\eeq
has a simple pole $s_{\rm pole}$ near $M^2$ in the second Riemann
sheet and for $s$ close to $s_{\rm pole}$ one gets
\beq
G(s)\sim\frac{i{\cal Z}}{s-s_{\rm pole}},
\eeq
where
\beq
{\cal Z}=\frac{1}{1-\Sigma_{\rm II}'(s_{\rm pole})} =1+\lambda^2
\Sigma^{(2)\prime}(M^2+i0^+)+{\rm O}(\lambda^4)
\eeq
and
\barr
s_{\rm pole}&=&M^2+\Sigma_{\rm II}(s_{\rm pole}) =M^2 + \lambda^2
\Sigma^{(2)}(M^2+i0^+)+{\rm O}(\lambda^4)
\nonumber\\
&=& M^2 - i \lambda^2 M \Gamma(M^2)
+{\rm O}(\lambda^4),
\earr
with
\beq
\Gamma(s)\equiv \frac{\mu^2}{32\pi M}\rho(s).
\eeq
The time evolution of the correlation function (\ref{eq:corr.func})
reads
\andy{time.evol}
\barr
\As(t)&\equiv& G(t,\bm p)=\int \frac{dE}{2\pi}\; e^{-iEt} G(p)\nonumber\\
&=&e^{-iE_p t}\int \frac{dE}{2\pi}\; e^{-iEt}
\frac{i}{E(2E_p+E)-\Sigma(M^2+E(2E_p+E))},
\nonumber \\
\label{eq:time.evol}
\earr
where $E_p=\sqrt{\bm p^2 + M^2}$ is the energy of the particle
$\Phi$. By introducing the rescaled time $t=\tilde t/\lambda^2$ and
energy $E=\lambda^2\tilde E$, Eq.~(\ref{eq:time.evol}) becomes
\beq
\As\left(\frac{\tilde t}{\lambda^2}\right)
=e^{-i\frac{E_p}{\lambda^2} \tilde t}
\int \frac{d\tilde E}{2\pi}\; e^{-i\tilde E\tilde t}
\frac{i}{\tilde E(2E_p+\lambda^2\tilde E)
-\frac{1}{\lambda^2}\Sigma(M^2+\lambda^2\tilde E(2E_p+\lambda^2\tilde
E))}
\eeq
and taking Van Hove's limit, we obtain
\andy{limit.evol}
\beq
\widetilde\As(\tilde t)=\lim_{\lambda\to 0}
e^{+i\frac{E_p}{\lambda^2} \tilde t}\As\left(\frac{\tilde
t}{\lambda^2}\right) =\int \frac{d\tilde E}{2\pi}\; e^{-i\tilde
E\tilde t}
\widetilde G(\tilde E)
\label{eq:limit.evol}
\eeq
with the limiting propagator
\barr
\widetilde G(\tilde E)&=&\lim_{\lambda\to 0}
\frac{i}{\tilde E(2E_p+\lambda^2\tilde E)
-\frac{1}{\lambda^2}\Sigma(M^2+\lambda^2\tilde E(2E_p+\lambda^2\tilde
E))}
\nonumber\\
&=& \frac{i}{2E_p \tilde E-\Sigma^{(2)}(M^2+i0^+)}\nonumber\\
&=& \frac{1}{2E_p}
\frac{i}{\tilde E+i\frac{M}{E_p}\frac{\Gamma}{2}}.
\earr
Therefore the time evolution (\ref{eq:limit.evol}) becomes
\beq
\widetilde\As(\tilde t)=\frac{1}{2 E_p}
\exp\left(-\frac{M}{E_p}\frac{\Gamma}{2}\tilde t\right)
=\frac{1}{2 E_p}\exp\left(-\frac{\tilde t}{2\tau_p} \right) ,
\eeq
and the particle decays exponentially with a mean lifetime
\beq
\tau_p=\frac{E_p}{M}\Gamma^{-1}=\frac{\Gamma^{-1}}{\sqrt{1-v^2}},
\eeq
which has the proper relativistic time dilatation factor. This result
is remarkably simple, for the model considered.

\setcounter{equation}{0}
\section{Conclusions and comments}
\label{sec-concom}
\andy{concom}

The time evolution obtained in Van Hove's limit is always purely
exponential: the quantum dynamics is governed by a master equation
and by dynamical semigroups. However, it is obvious that the very
procedure of time rescaling hides, in some sense, the problem of
state preparation.

In all the decay processes considered in this paper, an initial pure
state is considered at time``$t=0$." What is the meaning of $t=0$?
This question is often dismissed, in particular in quantum field
theory, where all ``relevant" physical quantities are constructed
from the $S$-matrix. There are however interesting examples in which
the issue of state preparation is discussed, both in the context of
semigroups \cite{ABohm} and scattering processes
\cite{POP}.
It is difficult not to wonder at the concept of initial time, in
particular when one considers fundamental processes like particle
creation in quantum field theory. Think again of the relativistic
model analyzed in the previous section, as well as of other examples
recently considered in the literature \cite{AGPP,BMT}. Any classical
picture of a decay process is necessarily mind-boggling. The
preparation of an initial wave function (or initial state of a
quantum field) is an inherently quantum mechanical process, certainly
not an easy one to conceive.

This problem is difficult to tackle. Nico Van Kampen, after
refereeing one of our papers, put forward the following interesting
and thought-provocative comment
\cite{vK}: ``As to your suggestion for preparing an initial pure
state, there is no objection to it, from the mathematical viewpoint.
But we are doing physics.  Your construction is of the same calibre
as the construction in statistical mechanics of those time reversed
states whose entropies increase.  The answer there too was that they
are permissible from the mathematical point of view, but that they
are tremendously improbable. Remember also the work by Wheeler and
Feynman, which argued that there can be no coherence in incoming
waves owing to the absorbtion property of the universe. My feeling is
that there is no real difficulty or paradox, but only the task to
formulate precisely what is intuitively clear."

We agree. Although an initial pure state like those considered in
this paper are {\em mathematically} easy to conceive, their {\em
physical} construction is prohibitive. Does this mean that
nonexponential decays in atomic or elementary particle physics are
extremely improbable to observe because ``unstable" quantum systems
are practically always created in some sort of mixed states, whose
statistical features justify time coarse-graining procedures like Van
Hove's? This is an interesting question, which goes to the very core
of the notion of irreversibility. A physical answer is needed.


%
%



\end{document}